\documentclass[]{spie}  
\usepackage[table]{xcolor}

\usepackage{amsmath,amsfonts,amssymb}
\usepackage{graphicx}
\usepackage[colorlinks=true, allcolors=blue]{hyperref}
\usepackage{textcomp}
\usepackage{gensymb}

\title{Refurbishment and Cryocooler Upgrade of the CPAPIR Wide-Field Near-Infrared Camera: Design, Integration, and Lessons Learned.}

\author[a,b]{Philippe Vall\'ee}
\affil[a]{Institut Trottier de Recherche sur les Exoplan\`etes (IREx), D\'epartement de Physique, Universit\'e de Montr\'eal, C.P. 6128, Succ. Centre-Ville, Montr\'eal, QC, H3C 3J7, Canada}
\affil[b]{Observatoire du Mont-M\'egantic, C.P. 6128, Succ. Centre-Ville, Montr\'eal, QC, H3C 3J7, Canada}
\author[a,b]{Jonathan St-Antoine}
\author[a,b]{Lison Malo}
\author[a,b]{Frédérique Baron}
\author[b]{Sylvie F Beaulieu}
\author[a,b]{Ren\'e Doyon}
\author[a,b]{\'Etienne Artigau}
\author[b]{Fidèle Robichaud}

\authorinfo{Further author information: (Send correspondence to P.V.)\\P.V.: E-mail: philippe.vallee@umontreal.ca, Telephone: 1 514 343 7118}

\graphicspath{{images/}}

\begin{document}

\maketitle

\begin{abstract}
The CAmera PAnoramique Proche Infrarouge ({CPAPIR\cite{artigau_cpapir_2004, httpswwwomm-astrocainstrumentcpapir_cpapir_2025 }}) is a wide-field near-infrared imager that has been operating at the Observatoire du Mont-Mégantic (OMM) since 2004. After more than two decades of service, the instrument required a major refurbishment due to detector degradation, including the loss of one quadrant, as well as operational constraints associated with its liquid-nitrogen ($LN_2$) cooling system.

To extend the instrument’s lifetime and reduce maintenance requirements, the detector was replaced and the cooling system upgraded through the integration of a Sunpower CryoTel DS30 cryocooler. This paper presents the thermal and mechanical redesign of the instrument, including finite element analysis (FEA), three-dimensional CAD modeling, hardware selection, and laboratory validation.

Particular attention is given to the challenges encountered during integration and testing, notably vibration mitigation and cryogenic system implementation. The resulting performance improvements and lessons learned from the upgrade are discussed. The upgraded instrument is scheduled for on-sky commissioning at the Observatoire du Mont-Mégantic in summer 2026.
\end{abstract}

\keywords{Wide field camera, infrared, cryogenic, Finite element analysis}


\section{SUMMARY}
This paper presents the upgrade of the CPAPIR camera. The selection and implementation of the cryocooler are discussed, along with the development of the thermal model and its comparison with experimental measurements.

Key challenges related to cabling, the glycol cooling loop, vibration, and entrance window condensation are also addressed.


\section{Introduction}
CPAPIR is a wide-field near-infrared camera that has been in operation at the Observatoire du Mont-Mégantic (OMM) since 2004, and was also deployed at CTIO in 2006. At OMM, it provides a field of view of approximately 0.5° × 0.5°, with a pixel scale of 0.88 arcseconds per pixel. The instrument is equipped with two cryogenic filter wheels, offering a total of 10 filters.

Thanks to its wide field of view, CPAPIR is particularly well suited for survey observations. It is also an effective instrument for studies of brown dwarf variability, as well as for observations of objects in young moving groups.

\begin{figure}[ht]
     \centering
     \includegraphics[width=17cm]{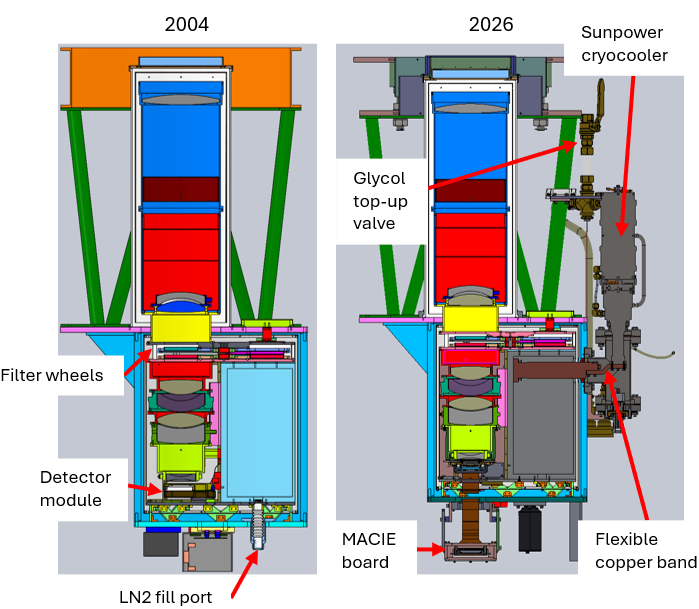}
     \caption{Cross-section view, CPAPIR 2004 (left) vs CPAPIR 2026 (right)}
     \label{fig: design_comparison}
 \end{figure}


\section{THE CRYOCOOLER CHALLENGE}
\subsection{Requirements}
The cryocooler had a set of competing requirements, summarized as follows:
\begin{itemize}
\item It must provide sufficient cooling power to bring the instrument, and in particular the infrared detector, close to its operating temperature of 80 K (target: below 100 K).
\item It must also enable a cooldown time short enough for practical operation (target: less than one week).
\item At the same time, it must be compact enough to be integrated into an existing instrument, while minimizing the number of components requiring modification or redesign.
\item The system must be as lightweight as possible in order to limit the torque applied to CPAPIR and to the telescope structure, as the instrument is directly mounted on the 1.6 m telescope.
\end{itemize}


\subsection{Cryocooler Selection}
Calculations based on a typical liquid-nitrogen hold time of $\sim$22 hours indicate a total thermal load of approximately 23 W for CPAPIR. This estimate guided the selection of the Sunpower CryoTel DS30 cryocooler, which provides 32 W of cooling power at 77 K with a heat rejection temperature of 23 °C.

It should be noted that the hold time was measured at the observatory, where the ambient temperature is typically lower than in laboratory conditions. To account for more demanding conditions, a conservative hold time of 16 hours—representative of summer operation—was also considered. Under these conditions, the estimated thermal load increases to approximately 32 W.

\subsection{Mechanical Design}
The transition from an $LN_2$ cooling system was primarily motivated by the need to simplify daily telescope operations and to improve the thermal stability of the instrument. The cryocooler is mounted parallel to the instrument (see Figure \ref{fig: design_comparison}), a configuration that minimizes the overall system envelope. However, this layout places the cold tip at a significant distance from the cold bench.

As a result, a temperature gradient develops between the cold tip and the cold bench, leading to an increase in the detector operating temperature, despite the cryocooler’s ability to provide more than the 23 W estimated from the $LN_2$ evaporation rate. To mitigate this effect, the design focused on:
\begin{itemize}
    \item maximizing the thermal conductance between the cold tip and the cold bench; 
    \item minimizing parasitic heat loads entering the instrument (see Section \ref{thermal_model} ). 
\end{itemize}
To achieve high thermal conductance, Oxygen-Free High Conductivity (OFHC) copper was selected for both the cold bus and the flexible thermal strap. The diameter of the cold bus was maximized within the constraints imposed by the largest flat surface that could be machined on the existing instrument enclosure. The flexible copper strap was designed to be as short and thick as possible to reduce thermal resistance, although this choice reduces its vibration isolation efficiency.

To further improve thermal conductivity at critical interfaces, a thin layer of vacuum grease (Apiezon N™) was applied. Finally, in order to minimize system mass and mechanical complexity, the vibration mitigation approach was intentionally kept simple. The cryocooler is rigidly mounted to the instrument, with vibration isolation provided primarily by the flexible copper strap (see Figure \ref{fig: design_comparison}).



\subsection{The Thermal Model}
\label{thermal_model}
\subsubsection{Minimization of Conductive Heat Loads}
Given that the CryoTel DS30 cannot be directly connected to the cold bench, and that its cooling capacity only marginally meets CPAPIR’s requirements, it became necessary to minimize parasitic heat loads entering the instrument.

\begin{figure}[ht]
     \centering
     \includegraphics[width=17cm]{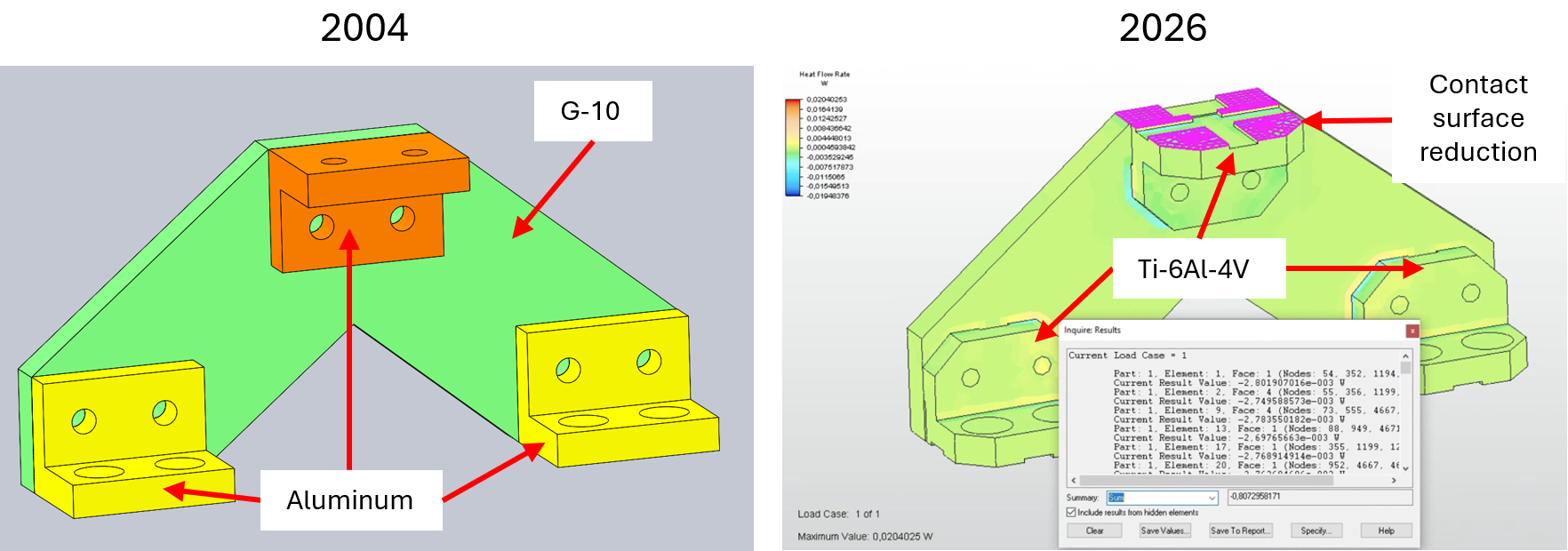}
     \caption{Model comparison and FEA results (right) for a main support, made of G-10/Aluminum/Ti-6AL-4V}
     \label{fig: main_support_comparison}
 \end{figure}

Three primary conductive heat paths were identified and modeled using finite element analysis (FEA) to evaluate potential reductions. The study investigated the replacement of aluminum components with titanium (Ti-6Al-4V), as well as a reduction of contact surface areas. The results are summarized in Table \ref{tab:heat_point}  and one case is illustrated in Figure \ref{fig: main_support_comparison}.

The modeling predicts a total reduction of 4.3 W. While modest in absolute terms, this corresponds to approximately 13\% of the total available cooling power (32 W), and is therefore significant.

 \begin{table}[h!]
    \centering
    
    \begin{tabular}{|p{3cm}|p{1.3cm}|p{1.3cm}|p{2cm}|p{2cm}|p{2cm}|} \hline 
    \rowcolor[HTML]{FFFFC7}
        \textbf{Component} & \centering \textbf{Before} \par \textbf{(W)} & \centering \textbf{After} \par \textbf{(W)} & \centering \textbf{Gain per} \par \textbf{ part (W)} & \centering \textbf{Number} \par \textbf{of parts} & \textbf{Total gain} \par \textbf{(W)} \\ \hline 
         G-10 main supports &  \centering 1.10 & \centering 0.81 &  \centering 0.29 & \centering 8 &  2.3 \\ \hline
     \rowcolor[HTML]{DAE8FC}    
         Warm radiation \par shield supports & \centering 0.56 & \centering 0.31 & \centering .25 & \centering 8 &  2.0 \\ \hline
        Filter wheel \par supports & \centering 0.258 & \centering  0.255 & \centering  0.003 & \centering 4 &  0.012  \\ \hline
    \end{tabular}
    \caption{ FEA results for conductive heat paths (before/after comparison).}
    Filter wheel supports were not modified, as the predicted thermal gain was negligible.
    \label{tab:heat_point}
\end{table}
 
\subsubsection{Minimization of Radiative Heat Loads}
 Radiative heat transfer becomes significant in the presence of large temperature gradients, as it scales with the fourth power of temperature according to the Stefan-Boltzmann law.
 
\begin{figure}[ht]
     \centering
     \includegraphics[width=17cm]{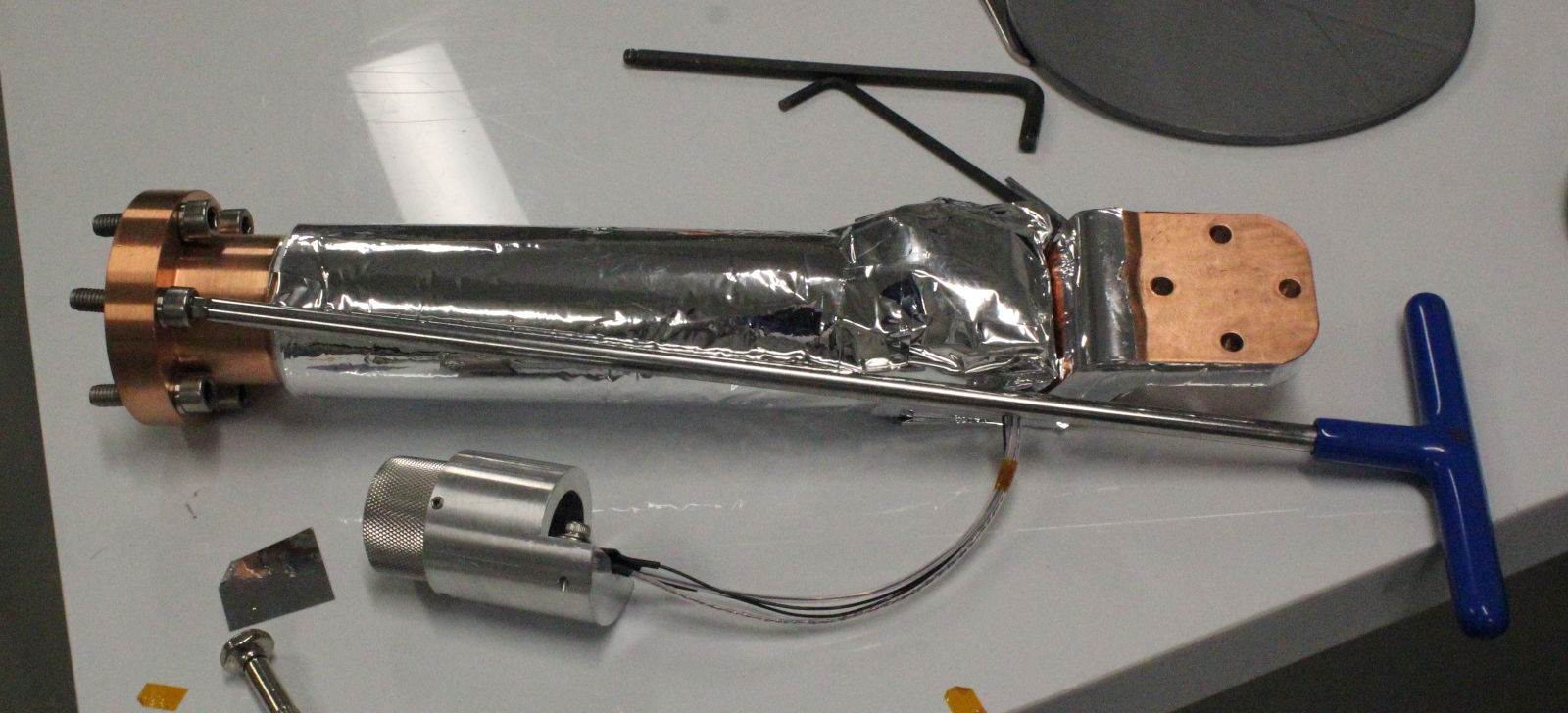}
     \caption{The cold bus wrapped with MLI and the screw driver used to install it}
     \label{fig: cold_bus_MLI}
 \end{figure}

 CPAPIR includes a floating thermal shield that reduces radiative heat loads. However, the cold bus and the cryocooler cold head are not enclosed within this shield and therefore remain exposed to thermal radiation. To mitigate these additional loads, multilayer insulation (MLI) was applied to the cold bus and within the interface tee connecting the cryocooler to the cryostat (see Figure \ref{fig: cold_bus_MLI}).
 
 Although multiple layers of MLI would provide improved performance, a single layer was used due to space constraints, minimizing the risk of thermal shorts between the cold bus and the instrument outer shell.
 
\subsubsection{Complete FEA Model}
To estimate the detector operating temperature and the time required to reach thermal equilibrium, a complete—yet simplified—FEA model of the instrument was developed.

Autodesk Fusion™ was used to evaluate multiple steady-state cases, while a transient model was implemented in ANSYS™ to estimate cooldown time and final operating temperatures. For both models, surface emissivities were adjusted to obtain heat flow values broadly consistent with those derived from the cryostat hold time, estimated at approximately 23 W under optimal conditions.

To reduce computational complexity and simulation time, several simplifications were introduced. The detector module was not included in either model. Similarly, the outer shell of the cryostat was not explicitly modeled, but instead represented by a boundary condition defined by ambient temperature and an effective emissivity. In the Fusion model, the lenses were not modeled, as they were not expected to affect steady-state temperatures.
\begin{table}[ht]
\centering
\begin{tabular}{|l|>{\centering\arraybackslash}p{0.1\textwidth}|>{\centering\arraybackslash}p{0.1\textwidth}|>{\centering\arraybackslash}p{0.15\textwidth}|}
\hline
\rowcolor[HTML]{FFFFC7} 
\textbf{Parameter} & \textbf{Fusion}    & \textbf{ANSYS} & \textbf{As measured} \\ \hline
Aluminum emissivity (\%) &  1 & 2 & N/A  \\ \hline
\rowcolor[HTML]{DAE8FC} 
Temperature at cold tip (K) & 77.2$^{1}$ & 70.9$^{2}$ & 72.6$^{3}$ \\ \hline
Temperature at detector mount (K) & 97.2 &  92.3 & 94.1 \\ \hline
\rowcolor[HTML]{DAE8FC} 
Heat flow through cold bus (W) & 27.6 & 20.5 & 31$^{4}$ \\ \hline
Cooldown time (days) & N/A & 3 & 4 to 5$^{5}$ \\ \hline
\end{tabular}
\caption{Comparison between thermal model prediction and measured values }
\label{tab:FEA_comparison}
\begin{flushleft}
Notes:

1. Temperature imposed in the model

2. Temperature after 3 simulated days

3. Temperature reached after 5 days

4. Calculated from measured temperature gradient along the cold bus

5. Experimental cooldown duration
\end {flushleft}
\end{table}
Table \ref{tab:FEA_comparison} compares the simulation results with laboratory measurements. The measured heat flow was estimated using {Fourier's law \cite{httpswwwmathworkscomhelpsimscapeugheat-conduction-through-iron-rodhtml_fourier_2026}}:
\begin{equation}
Q = k A \frac{T_2 - T_1}{L}
\end{equation}
where k is the thermal conductivity of copper (483 W·m\textsuperscript{-1}·K\textsuperscript{-1}), $T_1$ (74 K) and $T_2$ (85 K) are the temperatures measured at both ends of the cold bus, A (1.25 × 10\textsuperscript{-3} m\textsuperscript{2}) is its cross-sectional area, and L (0.212 m) is its length.

The ANSYS model appears to underestimate the heat flow, which likely explains why the predicted cooldown time is shorter than observed experimentally. From an operational standpoint, this discrepancy is not critical, as ambient temperatures at the observatory are significantly lower than in the laboratory and are expected to reduce cooldown time.

It was also observed that, even when using an emissivity twice as high, the ANSYS model still predicted a lower heat flow than the Fusion model. This discrepancy was not further investigated due to time constraints. Overall, the model predictions were considered satisfactory.

\begin{figure}[ht]
     \centering
     \includegraphics[width=17cm]{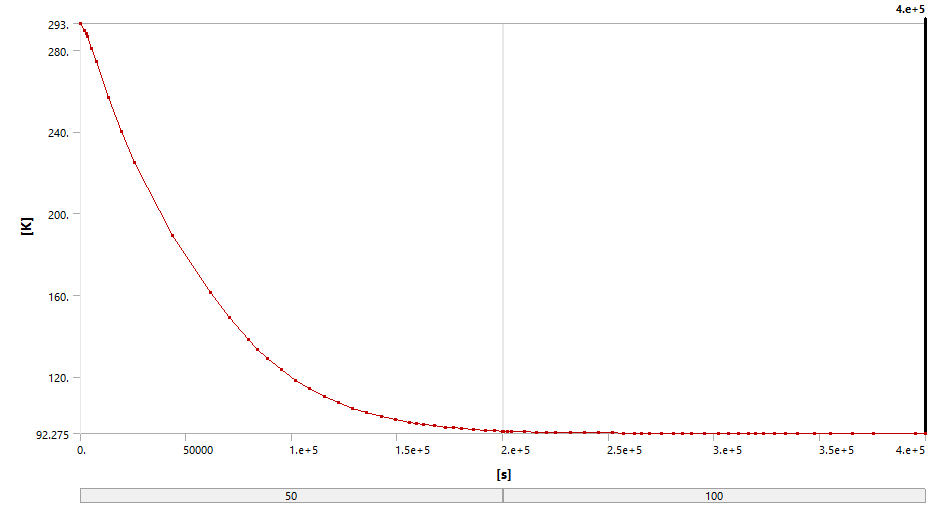}
     \caption{ANSYS cooling curve, showing the evolution of the temperature close to the detector}
     \label{fig: ANSYS_curve}
 \end{figure}

\begin{figure}[ht]
     \centering
     \includegraphics[width=17cm]{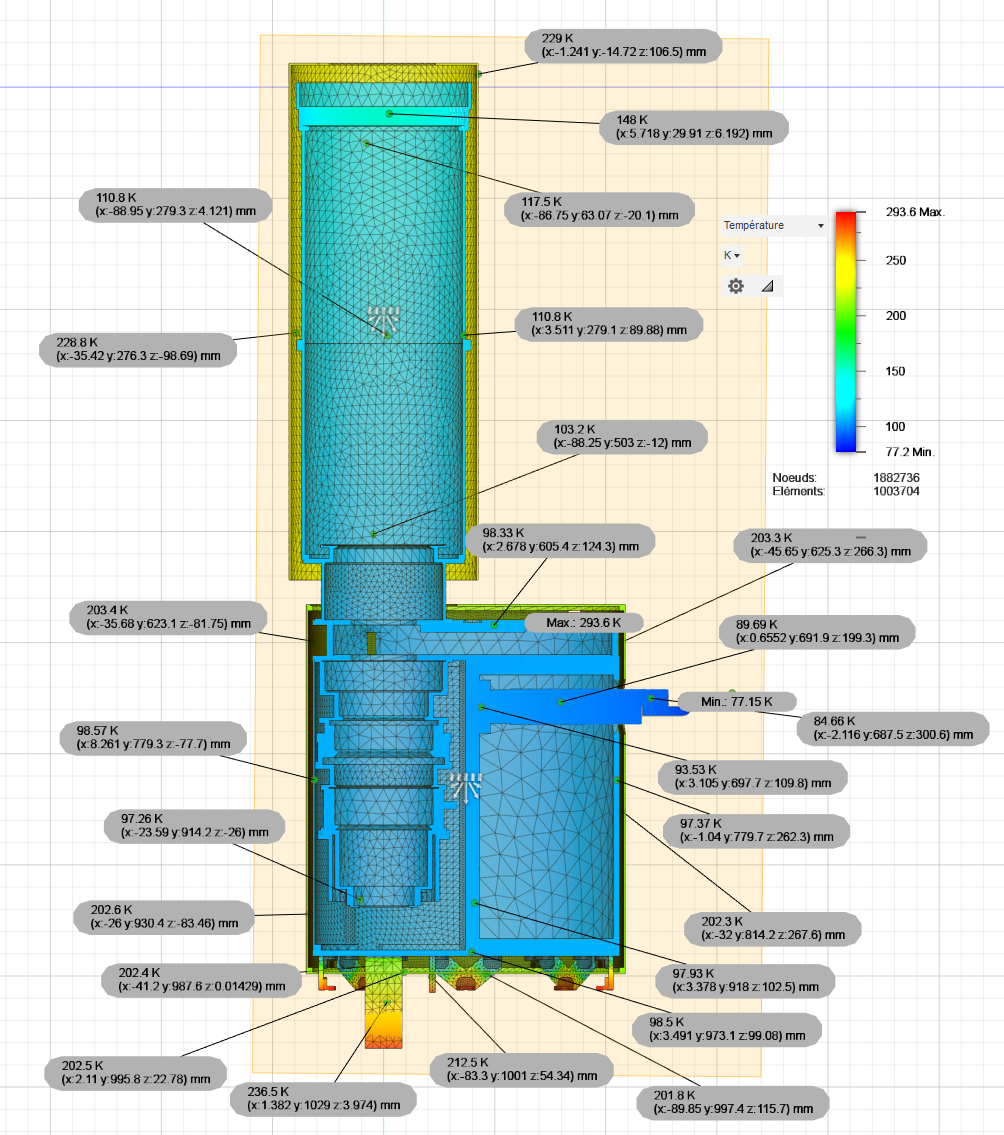}
     \caption{Fusion model, cross-section showing the temperature distribution at equilibrium}
     \label{fig: Fusion_V11}
 \end{figure}

\begin{figure}[ht]
     \centering
     \includegraphics[width=11.5cm]{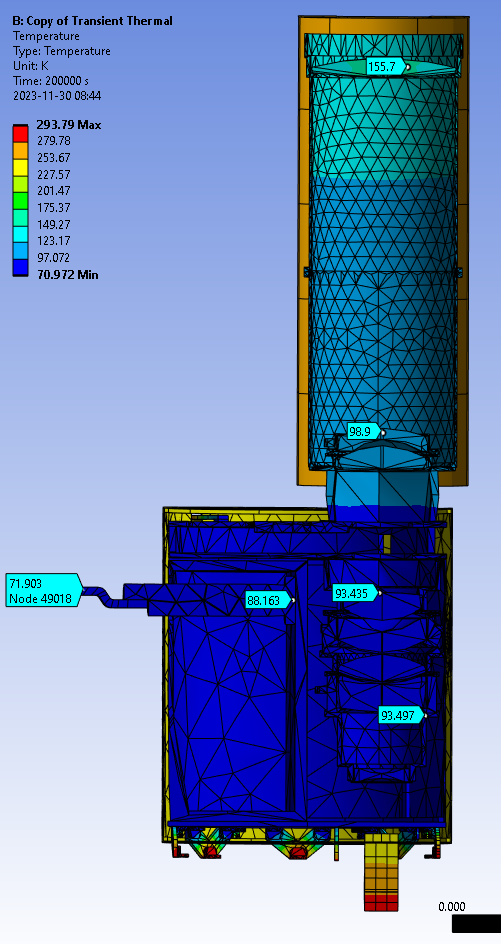}
     \caption{ANSYS model, cross-section showing the temperature distribution after 200 000 secondes}
     \label{fig: ANSYS_200 000}
 \end{figure}

Examples of temperature distributions obtained with Fusion and ANSYS are shown in Figures \ref{fig: Fusion_V11} and \ref{fig: ANSYS_200 000}, respectively, while the simulated cooldown curve from ANSYS is presented in Figure \ref{fig: ANSYS_curve}.

\begin{figure}[ht]
    \centering
    \begin{minipage}{0.49\textwidth}
    \centering
    \includegraphics[width=\textwidth]{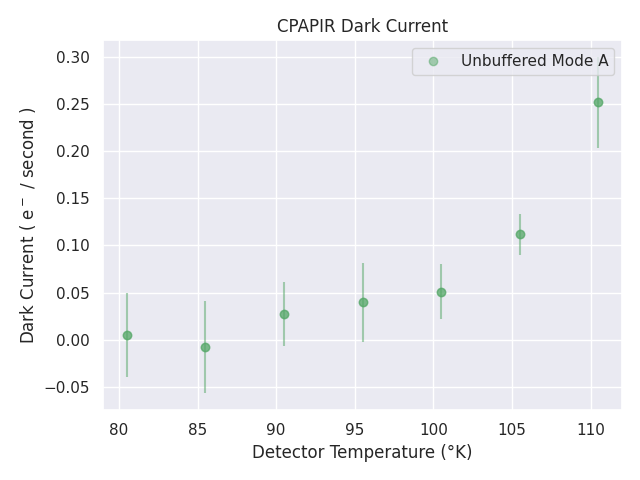}
    \caption{Dark current as a function of temperature}
    \label{fig: dark current}
    \end{minipage} \hfill
    \begin{minipage}{0.49\textwidth}
    \centering
     \includegraphics[width=\textwidth]{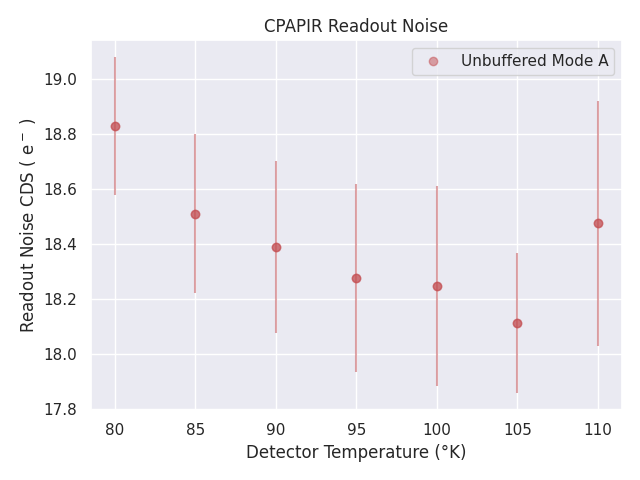}
     \caption{Read noise as a function of temperature}
     \label{fig: read noise}
     \end{minipage}
 \end{figure}

\newpage
Laboratory measurements currently indicate a detector operating temperature of approximately 95 K, although the detector has not yet been powered. A slight increase to $\sim$96 K is expected during operation. While this temperature is higher than initially targeted, prior testing of the detector in a separate cryostat demonstrated operation up to 100 K without significant degradation in dark current (Figure \ref{fig: dark current} ) or read noise (Figure \ref{fig: read noise} ).

\subsection{Glycol Cooling System}
The CryoTel DS30 cryocooler requires an external cooling system to dissipate the rejected heat. As the observatory does not provide a closed-loop cooling circuit, a dedicated system had to be implemented. For this purpose, a Lytron (now Boyd) Modular Cooling System (MCS, model MCS20G01BB1) was selected.

The MCS is a passive system consisting primarily of a pump, a radiator, hoses, and coolant. Although water is the recommended coolant, a glycol–water mixture was used to prevent damage from sub-freezing ambient temperatures.

The MCS is not designed to operate significantly below the rest of the cooling circuit, as coolant will drain back into the reservoir. Nevertheless, operation from a lower floor was successfully achieved by removing the original reservoir and installing a tee fitting with a fill valve at the highest point of the cooling loop. This modification allows the system to be filled and maintained from that location.

This high-point fill port proved essential during laboratory testing. Despite the use of quick-connect fittings, air was found to enter the loop during reconnection. Periodic refilling was therefore required to maintain proper operation; otherwise, the pump became excessively noisy due to the presence of air in the circuit.

The MCS provides a cooling capacity of up to 1300 W, assuming the coolant temperature rises to 25\degree C above ambient. The nominal flow rate is 4.9 L/min. The DS30 requires a flow rate of approximately 1 L/min at a maximum inlet temperature of 40\degree C and dissipates up to 480 W.

Laboratory measurements showed a coolant flow rate of 3 L/min and an inlet temperature of 31\degree C, under ambient conditions of approximately 22\degree C and using a 60 m-long cooling loop, representative of observatory operation. These results indicate that the system operates well within the required specifications for the DS30.

One remaining question concerns winter operation at the observatory, where ambient temperatures can drop below -15\degree C. Lower coolant temperatures may improve cryocooler performance by enhancing heat extraction. However, the coolant consists of a 60/40 glycol–water mixture, whose viscosity increases significantly at low temperatures—potentially by an order of magnitude between {27\degree C to -18\degree C\cite{httpscorecheminccomethylene-glycol-water-mixture-properties_viscosity_2026}}. The impact of this effect on system performance will be evaluated during future winter operations.

\subsection{Vibration Mitigation}
To minimize system mass and mechanical complexity, the vibration mitigation approach was intentionally kept simple. The cryocooler is rigidly mounted to the instrument, with vibration isolation provided primarily by the flexible copper thermal strap.

When the DS30 was first operated, significant vibrations were observed throughout the instrument. These vibrations appeared abnormally high and were independent of the vibration mitigation mode (VIBE ON or OFF).

The issue was ultimately traced to a fault in the controller microcode. After corrective action by the manufacturer, vibration levels were significantly reduced and appear acceptable, although independent quantitative measurements have not yet been performed.

Measurements obtained using the cryocooler controller indicate that vibration levels are comparable when the cryocooler is OFF and ON, except during startup, when the controller performs its tuning sequence.

This issue resulted in a delay of approximately five months in the project schedule.

\subsection{Cable Length Constraints}
CPAPIR is installed directly on the telescope, exposing it to low temperatures (down to -30\degree C) and high humidity ($>$90\%). To protect the electronics, it is preferable to locate them in the control room, where the temperature remains stable at approximately 20\degree C year-round.

The control room is therefore the optimal location for the DS30 controller and its power supply. However, this configuration requires cable lengths of approximately 20 m between the cryocooler and its controller.

Both long (20 m) and short (5 m) cable sets were fabricated and tested in the laboratory. The most critical connections were identified as the power supply lines and the accelerometer signal cable. The accelerometer cable consists of shielded 24 AWG wires, while the power cable uses 14 AWG conductors.

Tests showed that the DS30 operates similarly with both cable lengths. With the longer power cables, the minimum achievable temperature was approximately 1 K higher than with the shorter cables, likely due to resistive losses. During operation, the long power cables were observed to reach temperatures of approximately 47\degree C.

Based on these results, operation with long cables at the observatory is considered feasible.
\section{The detector challenge}
\subsection{Current Situation}
CPAPIR is currently equipped with a HAWAII-2 detector installed in 2004. One quadrant of the detector is defective, resulting in the loss of approximately one quarter of the field of view. This limitation reduces the efficiency of wide-field surveys and impacts the overall scientific return of the instrument. Replacement of the detector was therefore required.

This upgrade is not straightforward, as currently available detectors from Teledyne, such as the HAWAII-2RG, feature significantly different packaging and interface requirements. In addition, the existing CPAPIR data acquisition system is based on a Leach controller, with software running on a Sun workstation. Both systems are now obsolete and also required replacement.
\subsection{Solution}
A new detector was procured, with its packaging subcontracted to COMDEV (now Honeywell).
The data acquisition system has also been upgraded and now incorporates a {SIDECAR\cite{chen_sidecar_2014}} ASIC coupled with a MACIE interface board. The control software is based on developments previously implemented for the {NIRPS \cite{artigau_nirps_2024, wildi_nirps_2017}} and {SPIP\cite{challita_spip_2024}} instruments, and is now deployed on an Ubuntu-based system.

\subsection{Integration Constraints in a Confined Volume}
\begin{figure}[ht]
     \centering
     \includegraphics[width=17cm]{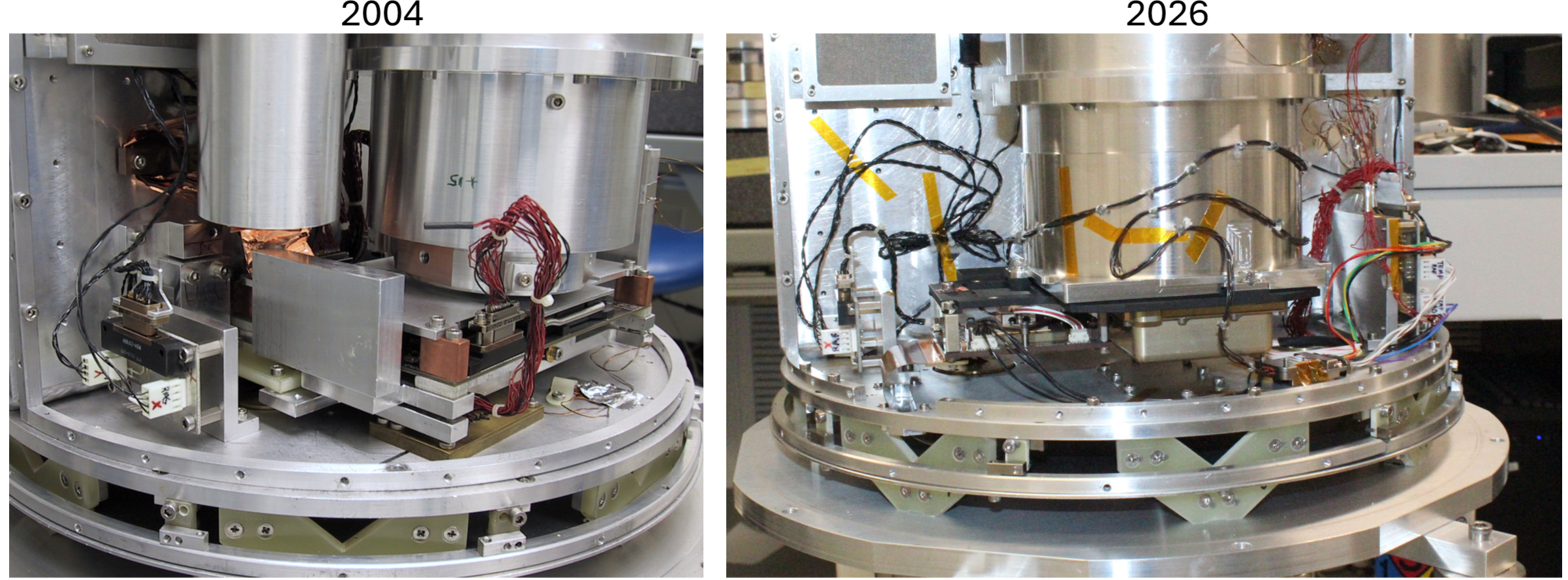}
     \caption{Comparison of the CPAPIR detector module in 2004 and 2026 configurations. Note that the 2004 module had extra lumps material to slow down the cooling by $LN_2$. They were removed in the 2026 version}
     \label{fig: detector_comparison}
 \end{figure}

The available space within the cryostat is limited, making the integration of the new detector, its SIDECAR, and the associated flex cable particularly challenging (see Figure \ref{fig: detector_comparison}).

The detector module was successfully installed in its as-built configuration; however, routing the flex cable proved more complex. While the instrument was still in operation at the observatory, several candidate routing configurations were prototyped using 3D printing to assess their feasibility.

\begin{figure}[ht]
     \centering
     \includegraphics[width=17cm]{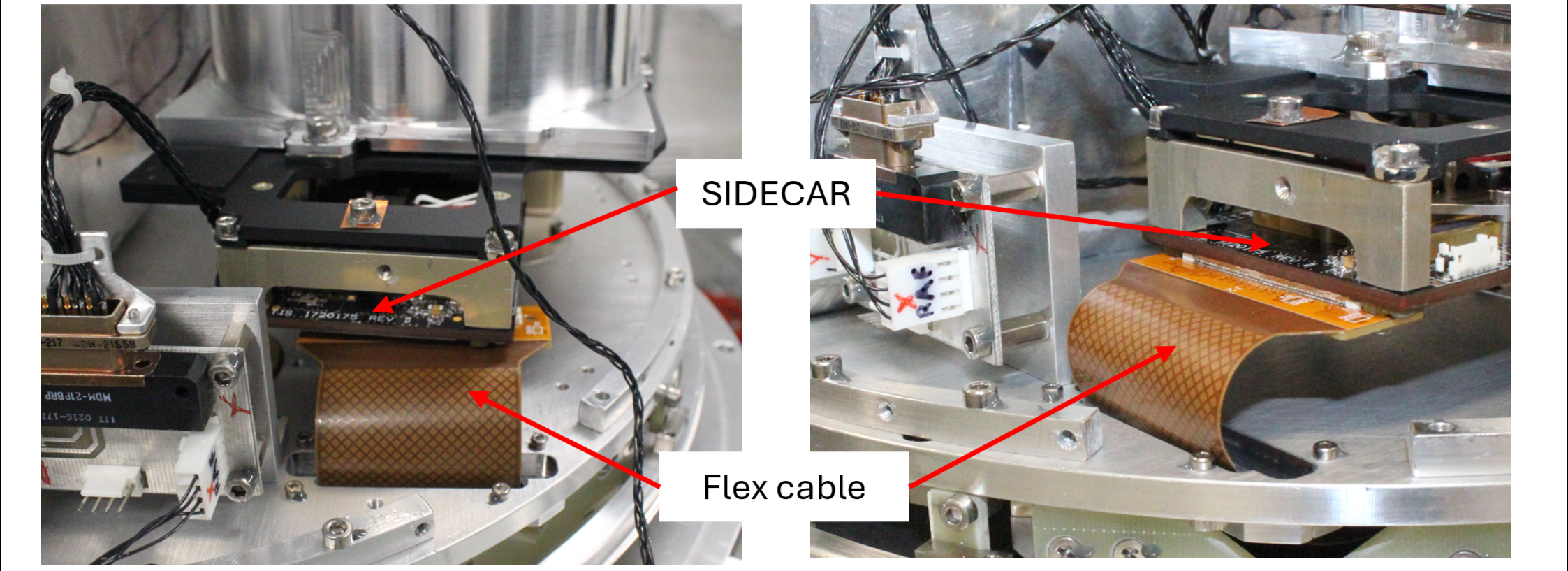}
     \caption{Misalignment between the flex cable and the SIDECAR (left). Extra holes were drilled in the camera barrel to correct the problem (right)}
     \label{fig: sidecar_misalign}
 \end{figure}

During final integration, it was found that the clocking of the camera barrel differed from the nominal geometry defined in the 3D model. This discrepancy resulted in a significant angular misalignment between the flex cable connector and the SIDECAR interface (see Figure \ref{fig: sidecar_misalign}). The misalignment was measured to be approximately 6.5\degree .

Corrective action required drilling additional mounting holes to realign the interface. Although this operation involved some risk, it was successfully carried out in the machine shop with the optical elements remaining installed in the camera.

This misalignment highlights the limitations of legacy mechanical references in refurbished instruments.

\subsection{Cabling and Feedthrough Challenges}
Cables used in infrared instruments must be properly thermally anchored to prevent parasitic heat loads. Without adequate thermal anchoring, warmer cable sections inside the cryostat can emit infrared radiation that may reach the detector and increase the background, thereby degrading instrument sensitivity.
In addition, cable interfaces must be light-tight to prevent external radiation from entering the cryostat. Airtight feedthroughs are also required to ensure long-term vacuum integrity.

\begin{figure}[ht]
     \centering
     \includegraphics[width=17cm]{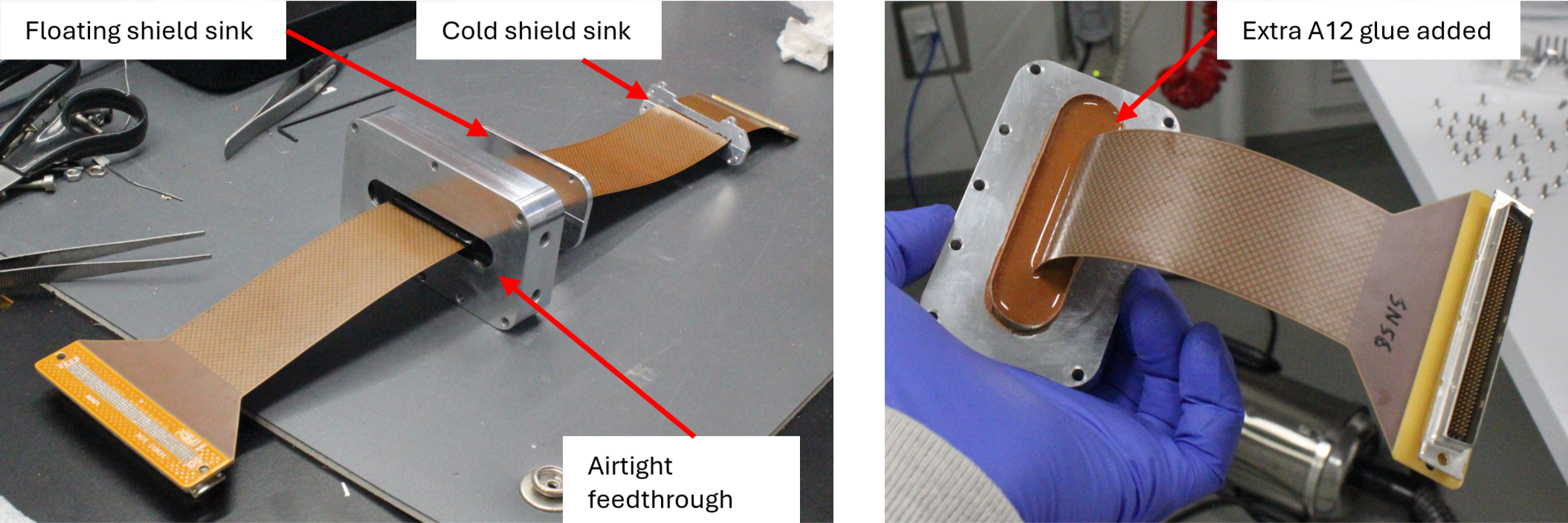}
     \caption{The flex cable, with its feed-through and anchoring glued on. Right picture shows the A12 glue that was added to fix the leak.}
     \label{fig: flex_cable}
 \end{figure}

To meet these requirements, two thermal anchors and a vacuum feedthrough were machined and bonded directly onto the detector cable provided by Teledyne (See Figure \ref{fig: flex_cable}). Dedicated fixtures were designed and fabricated to ensure precise bonding operations.

The initial feedthrough was assembled using Stycast 2850FT with Catalyst 23LV. Although this epoxy had been successfully used in previous vacuum-tight assemblies, the present implementation failed the leak test. The failure was attributed to the epoxy being significantly beyond its shelf life.

The leak was successfully corrected by applying an additional sealing layer using ResinLab Armstrong™ A-12 epoxy. This adhesive was selected for practical reasons: it was readily available and exhibits lower flow compared to Stycast, making it easier to control during application. PTFE components were used as temporary fixtures to contain the resin during curing, while a thin layer of vacuum grease was applied at critical interfaces to prevent unwanted flow.

Previous experience with the {NIRPS\cite{artigau_nirps_2024, wildi_nirps_2017}} instrument has shown that the cable between the {SIDECAR\cite{chen_sidecar_2014}} ASIC and the MACIE interface can be several meters long without performance degradation. However, implementing a long cable ($\sim$ 20 m) to relocate the MACIE electronics to the control room was not considered practical.

Instead, a protective enclosure for the MACIE board was designed and mounted directly on the cryostat, near the vacuum feedthrough. This solution allowed the use of the 15-inch flex cable provided by Teledyne.

One remaining open question concerns the operation of the MACIE electronics at low ambient temperatures, potentially well below freezing. This will be evaluated during winter commissioning at the observatory.

\subsection{Detector Flatness}
Inspection of the detector module, as assembled by COMDEV (now Honeywell), revealed a flatness of approximately 120\textmu m between the detector and its mounting surface. Such flatness deviations would have significantly impacted image quality and focus uniformity across the detector. It was deemed unacceptable.

The module was fully disassembled, and a detailed metrology analysis was performed on all components in order to identify and correct the source of the issue. One component required rework, and a precision shim was added, resulting in a final flatness of 28\textmu m, which meets the instrument requirements.

During this process, some detector bond wires were inadvertently damaged. The detector therefore had to be returned to the manufacturer for repair. This resulted in additional costs and a significant delay in the project schedule.

\section{Other improvements}
\subsection{New Entrance Window}
The original instrument window, installed in 2004, was made of fused silica, which has a relatively low thermal conductivity{($\sim$1.35 W·m\textsuperscript{-1}·K\textsuperscript{-1}) \cite{httpswwwmatwebcomsearchdatasheetaspxmatguid61499d82439f466eba1689158032dbacckck1_fused_2026}}. As a result, the center of the window would often cool below the dew point, leading to condensation and, in some cases, frost formation. This issue periodically rendered the instrument unusable.

To mitigate this effect, a flow of air was previously directed onto the window. However, this approach was inefficient, requiring a large airflow rate and a compressor that dissipated significant heat into the observatory environment.

A new window made of sapphire was therefore installed. Sapphire has a much higher thermal conductivity {($\sim$27.2 W·m\textsuperscript{-1}·K\textsuperscript{-1})\cite{httpswwwmatwebcomsearchdatasheetaspxmatguid8d86f08dbd424a6f872f9d0fe4976373_schott_nodate}} than fused silica, which should help maintain the window temperature above the dew point.

\begin{figure}[ht]
     \centering
     \includegraphics[width=17cm]{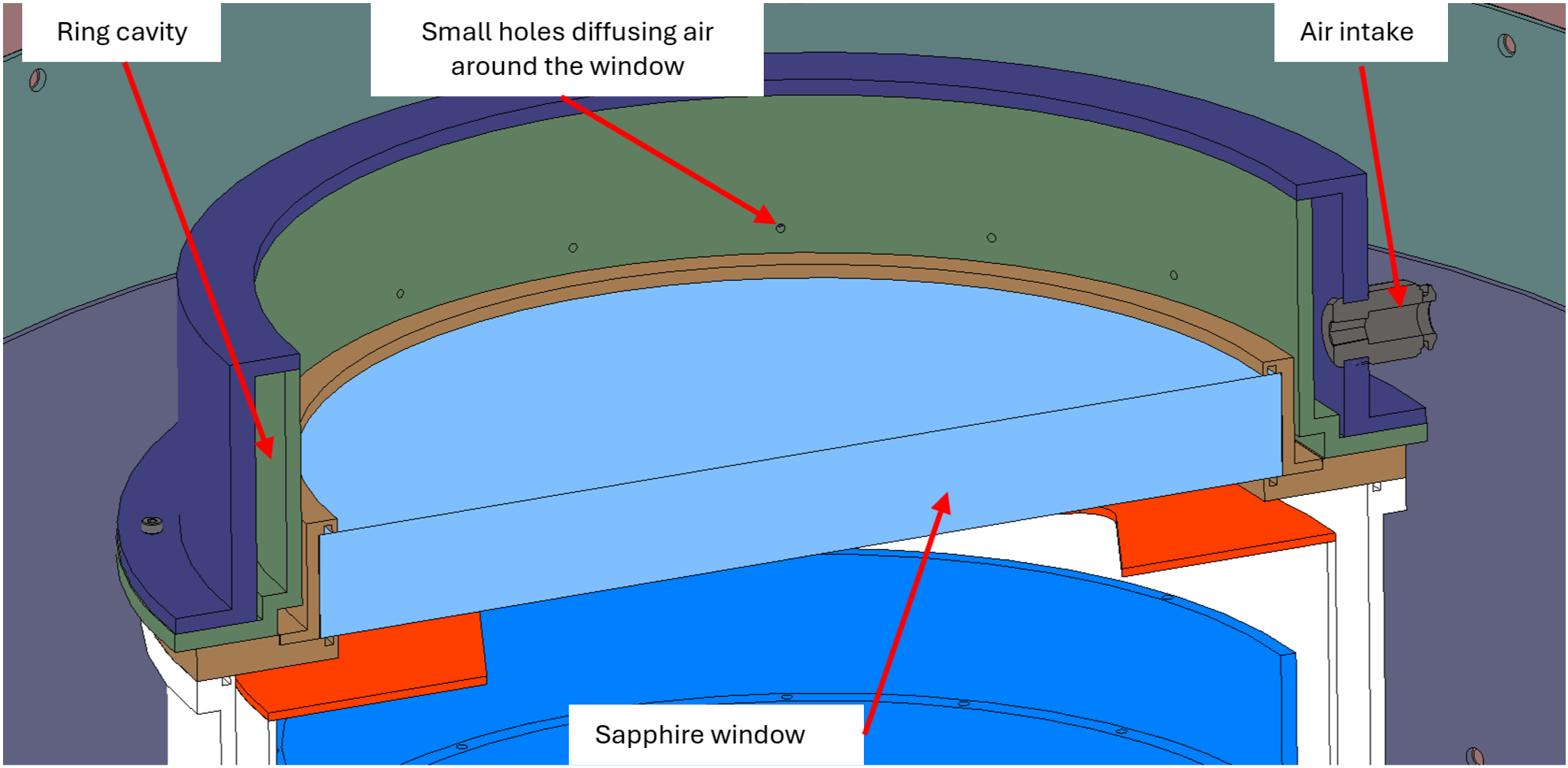}
     \caption{A cross-section view of the sapphire window and its hollow ring}
     \label{fig: sapphire_window}
 \end{figure}

In addition, a hollow ring was installed around the window to distribute airflow uniformly across its surface. This configuration is expected to create a localized pocket of dry air in front of the window while significantly reducing the required airflow. As a result, the compressor load and associated heat dissipation should be reduced.
The upgraded configuration is shown in Figure \ref{fig: sapphire_window}.

\subsection{Refurbishment of the Filter Wheel System}
CPAPIR is equipped with two cryogenic filter wheels, providing access to a total of 10 filters. After more than 20 years of operation, the system has demonstrated overall reliable performance, with the exception of an early failure caused by a loose screw that led to motor damage.

Cryogenic mechanisms must be lubricated using dry lubricants; molybdenum disulfide (Mo$S_2$) powder is typically used for this purpose. After extended operation, inspection and maintenance of both filter wheels became necessary.

The assemblies were fully disassembled, and the sapphire balls used in the thrust bearings were replaced. The steel ball bearings were also replaced with new, degreased units and lubricated with Mo$S_2$. The same lubricant was applied to surfaces in contact with the sapphire balls.

Following reassembly, the filter wheels were successfully tested at both room temperature and cryogenic temperatures. This refurbishment is expected to ensure long-term reliability of the filter wheel mechanism under cryogenic operating conditions.
\subsection{Epoxy Staking of Fasteners}
The use of epoxy to secure fasteners against loosening due to thermal cycling and vibration is a common practice in aerospace applications.

During the original assembly of CPAPIR, no thread-locking or bonding agent was applied to the screws. This omission led to a filter wheel failure when a fastener became loose. Following this incident, the affected screws were bonded to prevent recurrence.

Although no additional failures were observed elsewhere, a decision was made to proactively secure all accessible fasteners using epoxy staking.

A small amount of 3M Scotch-Weld™ Epoxy Adhesive 2216 was applied to each fastener, forming a bond between the screw head and the adjacent fixed surface. This measure is expected to improve long-term mechanical reliability of the instrument.




\section{Conclusions}
Upcoming tests at the observatory are promising; however, several challenges remain.

To minimize system mass and mechanical complexity, the vibration mitigation approach was intentionally kept simple. In retrospect, a more compliant thermal link would likely have been preferable. The use of copper braid instead of copper foil should have been considered, although the resulting reduction in thermal conductivity would likely have increased the instrument operating temperature. To compensate for this loss, a more optimized cold bus design could have been implemented, particularly given the available volume previously occupied by the $LN_2$ tank.

Further vibration mitigation could also be achieved by mounting the cryocooler using a more compliant interface, similar to approaches adopted in instruments such as {GIRMOS\cite{dutt_final_2024}}.

The current solution to mitigate window condensation may prove insufficient under harsh environmental conditions. An active control system incorporating temperature and humidity measurements, combined with controlled heating of the window assembly, could be considered for future improvements.

Some issues encountered during integration—such as the controller microcode malfunction and the detector flatness problem—could likely have been identified earlier through more comprehensive acceptance testing upon component delivery.

Overall, the upgrade demonstrates that extending the lifetime of a legacy infrared instrument through targeted refurbishment is both feasible and effective. CPAPIR is expected to remain a valuable scientific instrument at OMM in the coming years by enabling continued wide-field near-infrared survey capabilities at OMM, particularly for time-domain and low-temperature stellar population studies.

\bibliography{reference_vallee_6} 
\bibliographystyle{spiebib} 
Disclosure : the text of this article was reviewed by Co-Pilot for orthography, grammar, and style.
\end{document}